\documentstyle[prb,preprint,aps]{revtex}
\draft
\input{psfig}
\psfigurepath{FIG/}

\def\AmS{{\protect\the\textfont2
        A\kern-.1667em\lower.5ex\hbox{M}\kern-.125emS}}

\makeatletter
\def\thepage{1-\@arabic\c@page}
\def\@pnumwidth{2em}
\makeatother
\begin{document}

\title{Current carrying capacity of carbon nanotubes}
\author{M. P. Anantram}
\address{NASA Ames Research Center, Mail Stop T27A-1, Moffett Field, 
CA, USA 94035-1000  }
\maketitle

\begin{abstract}
The current carrying capacity of ballistic electrons in carbon
nanotubes that are coupled to ideal contacts is analyzed. At small
applied voltages, where electrons are injected only into crossing
subbands, the differential conductance is $4e^2/h$. At applied voltages
larger than $\Delta E_{NC}/2e$ ($\Delta E_{NC}$ is the energy level
spacing of first non crossing subbands), electrons are injected into
non crossing subbands.
The contribution of these electrons to current is determined by the
competing processes of Bragg reflection and Zener type inter subband
tunneling. In small diameter nanotubes, Bragg reflection dominates,
and the maximum differential conductance is comparable to $4e^2/h$.
Inter subband Zener tunneling can be non negligible as the nanotube
diameter increases because $\Delta E_{NC}$ is inversely proportional to
the diameter. As a result, with increasing nanotube diameter, the 
differential conductance becomes larger than $4e^2/h$, though not 
comparable to the large number of subbands into which electrons are 
injected from the contacts. These results may be relevant to recent 
experiments in large diameter multi-wall nanotubes that observed 
conductances larger than $4e^2/h$.
\end{abstract}

\vspace{0.4in}

 Journal Refereence: Phys. Rev. B, vol. 62, p. 4837, (2000)

\pagebreak

{\it Introduction:}
Most experimental~\cite{Tans97,Cobden98,Soh99} and theoretical work of
electron transport in individual nanotubes deals with single wall 
nanotubes (SWNT). In these experiments, the spacing between subbands
is typically larger than the applied voltage and thermal energy $kT$.
Recent experiments on multi-wall nanotubes 
(MWNT)~\cite{Frank98,Schonenberger99,Pablo99} are fundamentally 
different in that the subband spacing is comparable to the applied 
voltage and only a few times larger than the room temperature kT. It is
further believed that transport in these experiments primarily takes 
place along individual layers, with little inter-layer coupling. 
From the view point of molecular electronics, the relatively small low
bias resistance of 500 $\Omega$ in multi-wall nanotube wires reported
in Ref. \onlinecite{Pablo99} is very promising. 
In addition, Ref. \onlinecite{Frank98} found that the increase in
differential conductance with applied voltage was not commensurate with
the increase in number of subbands in large diameter nanotubes. On the
theoretical side, Ref. \onlinecite{Lin97} found long tails in the
screening properties of metallic nanotubes. 
Motivated by the above work, we study the current carrying capacity
of electrons injected into a nanotube by including the non crossing
subbands.

{\it Central idea and basic processes involved:}
The central idea of this paper is that an applied bias across the 
nanotube results in a {\it transport bottle neck} due to Bragg 
reflection. This results in a smaller than expected increase in 
differential conductance with increase in applied bias.

In a defect free nanotube connected to ideal contacts, there are three
possibilities for an electron injected from the left contact (Fig. 1):
(i) Direct transmission, where an electron is transmitted in the
injected subband (solid line of Fig. 1),
(ii) Bragg reflection, which occurs when the wavevector (k) of an
injected electron evolves to a value where the velocity in subband $n$,
$v_n(k) = \frac{1}{\hbar} \frac{dE_n(k)}{dk} =0$ (subband extrema).  
In Fig. 1, an electron injected from the left contact into a non 
crossing subband undergoes Bragg reflection at the location of the 
arrow (dotted line), and
(iii) Inter subband Zener type tunneling, which involves tunneling 
between subbands induced by an electric field.
The spacing between non crossing subbands ($\Delta E_{NC}$ of Fig. 1)
decreases inversely with increase in nanotube diameter ($D$), 
$\Delta E_{NC} \propto 1/D$. So, we surmise that Zener tunneling 
becomes increasingly important in determining the I-V curve with 
increase in nanotube diameter.
The relative importance of these three phenomena depends on the energy,
potential profile, and nanotube diameter, as discussed in this paper.

{\it Method:}
The current is computed using the Landauer-Buttiker formula,
\begin{eqnarray}
I = \int \: dE \: T(E) \: [f_L(E) - f_R(E) ] \mbox{ ,} \label{eq:LB}
\end{eqnarray}
where, $T(E)$ is the total transmission (single particle transmission
probability summed over all subbands), and $f_L(E)$ and $f_R(E)$ are
the Fermi factors in the left ($\mu_L$) and right ($\mu_R$) contacts 
respectively. We calculate the total transmission within the context of
the pi-orbital approximation, where the hopping parameter is assumed to
be 3.1 V. The method to calculate the total transmission is the same as
in Ref.  \onlinecite{Anantram98}. The calculation of the I-V curve 
requires the potential drop across the nanotube. The potential drop 
should in principle be determined by the self-consistent solution of 
Poisson's equation and the non equilibrium electron density. This is a
difficult problem for nanostructures. So, to convey the essential 
physics illustrating the role of Zener tunneling, we assume analytical
profiles to simulate different values of the electric field, as 
discussed below. Finally, the nanotube is assumed to be coupled to 
ideal contacts, which means semi-infinite nanotube leads. So a work
function mismatch between the nanotube and real contacts, and the 
accompanying electrostatics is neglected.

{\it Results:}
The total transmission (and hence current) is determined by two 
physical parameters: 

\noindent
{\bf (i)} $\Delta E_{NC}$, the energy level spacing between the first
non crossing subbands (Fig. 1). 
$\Delta E_{NC}$ depends on diameter, and for a (N,N) nanotube,
\begin{eqnarray}
\Delta E_{NC} = 2 t_0 sin(\frac{\pi}{N}) \mbox{ ,}
\end{eqnarray}
where, $t_0$ is the hopping parameter between nearest neighbor carbon 
atoms. At applied voltages smaller than $\Delta E_{NC}/2e$, there is 
net injection of electrons only into the two crossing subbands.
When the applied voltage is larger than $\Delta E_{NC}/2e$, electrons
are injected into the non crossing subbands. The electrons injected 
into the non crossing subbands can in principle contribute to current
only if final states into which they can be transmitted are available.
As can be seen from Fig. 1, when $V_a \geq \Delta E_{NC}/e$, electrons
incident into the first non crossing subband below the band center
(in the left contact) can tunnel into states of the first non crossing
subband above the band center (in the right contact). $\Delta E_{NC}/e$
is the equivalent of the barrier height for this tunneling process.

\noindent
{\bf (ii)} The length over which an applied bias drops. This 
corresponds to the barrier length [the distance across which an 
electron should tunnel to reach a right moving state (Fig. 1)] for 
electrons injected into the first non crossing subband.

As $\Delta E_{NC}$ varies with diameter, we consider nanotubes with
diameters varying from 6.8$\AA$ to 27.2 $\AA$. They are the (5,5) 
[3.64eV], (10,10) [1.92eV], (13,13) [1.48eV],
(16,16) [1.22 eV] and (20,20) [0.98 eV] nanotubes ($\Delta E_{NC}$ is
given in the square brackets). The barrier length (and so the electric
field) is varied by considering the potential to drop linearly in 
sections that are 10$\AA$, 30$\AA$ and 60$\AA$ long. 
We have also calculated the effect of Zener tunneling by taking the
applied potential ($V_a$) to drop across the nanotube as,
\begin{eqnarray}
V(x) = \frac{V_a}{2} \{ 1 +
\frac{ 1+e^{  \frac{L_t}{L_{sc}} } }{ e^{\frac{L_t}{L_{sc}}} -
e^{-\frac{L_t}{L_{sc}}} } \; e^{-\frac{x}{L_{sc}}} -
\frac{ 1+e^{ -\frac{L_t}{L_{sc}} } }{ e^{\frac{L_t}{L_{sc}}} -
e^{-\frac{L_t}{L_{sc}}} } \; e^{ \frac{x}{L_{sc}}}  \}
\mbox{ ,} \label{eq:V}
\end{eqnarray}
where, $L_t$ is the length of the nanotube, and a typical value of
$L_t$=2500$\AA$. $L_{sc}$ is a parameter that determines the nature of
the voltage drop and $x$ is the nanotube axis. $L_{sc} > L$ corresponds
to a linear voltage drop. $L_{sc} < L$ corresponds to a scenario with
large potential drops near the left and right ends, and a flat 
potential in between. As a result, for an applied voltage, the
maximum electric field is smaller when Eq. (\ref{eq:V}) is used
instead of a linear potential drop. As a result of this the Zener
tunneling strength is smaller than in the linear case, and the total
transmission versus energy has a different form. There is however
no significant qualitative change of the main points discussed in the
paper when Eq. (\ref{eq:V}) is used.

The results for the 60$\AA$ case is discussed first. At applied 
voltages smaller than $\Delta E_{NC}/2e$, there is net injection of
electrons only into the two crossing subbands. As a result, the I-V 
curve is linear, with the differential conductance equal to $4e^2/h$
[Fig. 2]. This is true more or less independent of the distance over
which the voltage drops. For $V_a>\Delta E_{NC}/2e$, electrons are
injected into the higher subbands. Yet the maximum differential 
conductance in Fig. 2(a) is approximately $4e^2/h$. This is because 
electrons injected in the non crossing subbands are primarily 
reflected, and so do not carry an appreciable current.
To see this more clearly, consider the case of $V_a=2.5$V. Here, 
electrons are injected from the left into twenty subbands.
In Fig. 3, we show that for the 60$\AA$ case, all non crossing subbands
(solid and dotted lines) are almost fully Bragg reflected, and the 
crossing subbands are fully transmitted. Hence, the maximum differential
conductance in Fig. 2(a) is approximately equal to $4e^2/h$.
Alternately, electrons incident in the {\it non-crossing} subbands have
to traverse a spatial region with only the {\it crossing} subbands, 
before tunneling into the right contact. Hence, in the absence of 
significant inter subband tunneling, they are reflected. The above 
picture changes at voltages above 3.1V, which corresponds to a subband
extrema of the crossing subbands. When $V_a > 3.1$V, there is almost
no increase in current with applied voltage, in the voltage regime 
considered. This regime is explained by using $V_a=3.5V$ [Fig. 4]. When
$V_a=3.5$V, electrons are injected into thirty five subbands from the
left contact at $E=-V_a$. However, in a small energy range near $E=0$
and $-V_a$, the total transmission is approximately one. This is
because electrons injected into one of the two crossing subbands near
E=0 are Bragg reflected (at E=0, only one of the two crossing subbands
has a right moving state in the right contact). Similarly, 
near $E=\mu_L$, only one crossing subband has a right moving state at
the left contact. In between these energy windows with unity 
total transmission, both crossing subbands are transmitted. The
electrons injected into the non crossing subbands are almost fully 
Bragg reflected as discussed in the case of
$V_a=2.5$V. Upon increasing the applied voltage, the energy ranges
where a single crossing subband carries current broadens, and the
central energy range in between where both crossing subbands carry
current [Fig. 3] becomes narrower. The current, which is approximately
the integral of the area under the curve between E=0 and $-V_a$ does 
not increase much with further applied voltage as shown in Fig. 2.
The explanation for the other nanotubes in Fig. 2(a) is no different
except that the larger value of the barrier height $\Delta E_{NC}$
makes Bragg reflection only more important when compared to the
(20,20) nanotube.

The I-V characteristics in Fig. 2(a) were primarily determined by the
crossing subbands because all other subbands are almost completely
Bragg reflected.
This picture changes when inter subband Zener tunneling or defect
induced inter subband scattering are non negligible. To elucidate 
their effect on the I-V curve, we study their effects independently.

Zener tunneling, in principle begins to occur when
$V_a=\Delta E_{NC}/e$ [Fig. 1]. At this voltage, electrons
incident in the first non crossing subband below E=0 (in the left
contact) are able to tunnel into states of the first non crossing
subband above E=0 (in the right contact), as shown in 
Fig. 1. $\Delta E_{NC}/e$, which is the barrier height decreases
inversely with increase in nanotube diameter [the diameter is directly
proportional to N for a (N,N) nanotube], $\Delta E_{NC} \propto 1/N$ 
[Eqn. (1)]. So, we expect Zener tunneling to become more important with
increase in nanotube diameter. Figs. 2 (b) and (c) shows that the
calculated I-V deviates significantly from Fig. 2(a) as a result of
inter subband Zener tunneling. The main points are that Zener tunneling
and hence deviation of current from Fig. 2(a) increases,
(i) at smaller applied voltages, as the nanotube diameter increases
and the corresponding barrier height decreases [Figs. 2(b) and (c)],
and (ii) with increasing electric field [Figs. 2(a)-(c)] as the 
distance over which an electron has to tunnel is smaller.
A plot of the total transmission versus energy throws further
light on the physics involved. When the bias drops over 30$\AA$,
the opening and closing of a transmission window due to the first non
crossing subband (solid line marked NC1) is at energies of about 0.5eV
and 2.0eV respectively. These energies correspond to $\Delta E_{NC}/2$
below (0.5eV) and above (2.0eV) the nanotube band centers near the left
and right contacts respectively. In the case of L=10$\AA$, the
opening and closing of a second transmission window due to the second
non crossing subband (dotted line marked NC2) at energies of about 1eV
and 1.5eV respectively is seen. Also, the transmission probability
of NC1 is larger in comparison to the 30$\AA$ case because of the 
smaller barrier length. While the differential conductance at 
$V_a=2.5$V is not comparable to the twenty injected subbands, the 
contribution to current due to Zener tunneling cannot be neglected.

Frank et. al reported a constant conductance for applied voltages
smaller than an estimate of $\Delta E_{NC}/e$ for their large diameter
nanotubes, and a modest increase in conductance with applied bias
for larger applied voltages.~\cite{Frank98} This increase in 
conductance did not reflect the large increase in the number of 
subbands with bias. The transport bottle neck discussed in this paper
offers a possible physical mechanism that qualitatively explains the
small increase in conductance with applied voltage in Ref.
\onlinecite{Frank98}.

Finally, we discuss the role of defect assisted inter subband scattering.
From a physical view point, Bragg reflection is weakened because defect
scattering produces a non zero probability for an electron incident in
a non crossing subband to reach the right contact, by scattering into
right moving states of other subbands. To 
model defects, we follow section IIIA of Ref. \onlinecite{Anantram98},
where the on-site potential is varied randomly. We consider a 2500$\AA$
long nanotube section with defects, and the applied voltage drops 
linearly. So, Zener tunneling is not important here, and all inter 
subband tunneling is defect induced. The I-V curve is shown for two 
different strengths of defect scattering in Fig. 5. The numbers in the
legend correspond to $\epsilon_{random}$ of Ref. 
\onlinecite{Anantram98}, and a larger value corresponds to larger 
defect scattering. In Fig. 5, at  small voltages, only the crossing 
subbands determine the physics. Hence, the reflection of electrons
in the crossing subbands causes a diminished current and differential
conductance in comparison to the defect free case. At higher applied 
voltages, the differential conductance is larger than in the defect 
free case because the non crossing subbands are partially transmitted.
Transmission of electrons incident in the non crossing subbands is 
illustrated in the inset of Fig. 5, which is a plot of total 
transmission versus energy with and without defect scattering
($\epsilon_{random}=0.25 eV$) at $V_a=4V$, for $-V_a<E<0$. 
Defect scattering enhances the total transmission near $E=0$ and
$-V_a$ to values larger the defect free case. Thus showing that 
electrons injected in the crossing subbands are transmitted in
non crossing subbands at the right end (inset of Fig. 5).

While both Zener tunneling and defect scattering enhance the differential
conductance at large applied voltages, the following features
differentiate them. At biases smaller than $\Delta E_{NC}/e$, defects 
cause a reduction in current in comparison to the Zener tunneling case,
which continues to yield a conductance of $4e^2/h$. At biases larger 
than $\Delta E_{NC}/e$ electrons are
injected into many subbands. The differential conductance in
the case of Zener tunneling is larger than $4e^2/h$. 
Defect scattering alone, on the other hand, produces a differential 
conductance that is smaller than $4e^2/h$. 
This is because electrons incident in the {\it non-crossing}
subbands have to traverse a spatial region where only the {\it crossing} 
subbands are present, before being transmitted to the right contact.

In conclusion, we have investigated the current carrying capacity of 
carbon nanotubes by including transport through non crossing subbands.
This study considered ballistic transport and neglected 
electron-phonon interaction.~\cite{Yao00} We 
showed that due to the unique band structure of carbon nanotubes, 
Bragg reflection of electrons incident in the non crossing subbands is
an important mechanism for the reduction of differential conductance.
The differential conductance of nanotubes will be diameter dependent
from purely ballistic processes due to competition between Bragg 
reflection and Zener type inter subband tunneling. The importance of
Zener tunneling was studied by varying both the nanotube diameter
and the length over which the voltage drops. The barrier height for 
Zener tunneling is equal to the inter subband energy level spacing 
$\Delta E_{NC}$ of Fig. 1, which decreases inversely with increase in
nanotube diameter.
As a consequence, for small diameter nanotubes, the differential 
conductance cannot be larger than $4e^2/h$ for voltages smaller than 
3.1V, and is close to zero at larger applied voltages. Zener tunneling
becomes more important with increasing nanotube diameter because
$\Delta E_{NC} \propto$ 1/Diameter. Also, Zener tunneling is stronger 
when the voltage drops across a smaller length. We show for increasing
nanotube diameter, the total transmission is not negligible in certain 
energy windows, where the non crossing subbands carry current [Fig. 3].
As a result, the differential conductance at biases larger than 
$\Delta E_{NC}/e$ is larger than $4e^2/h$ [Fig. 2]. It should be
emphasized that the differential conductance is however not comparable
to the large number of subbands into which electrons are injected from
the contacts. The role of defect scattering in the absence of
Zener tunneling is also discussed. It is shown that at biases smaller
than 3.1V, defect scattering leads to a differential conductance that
is smaller than $4e^2/h$. For biases larger than 3.1 V, defects 
increase the differential conductance when compared to the defect free
case. 

Useful discussions with Zhen Yao, Cees Dekker, T. R. Govindan, Adrian
Batchold and Walt de Heer are acknowledged. I would like to thank
Bryan Biegel for correcting an earlier version of the manuscript and 
discussions.

\pagebreak

\noindent
{\bf Figure Captions:}

Fig. 1: Each rectangular box is a plot of energy versus wavevector,
with the subband bottom equal to the electrostatic potential. Only a
few subbands are shown for the sake of clarity. The three processes
shown are direct transmission (solid line), Bragg reflection (dotted
line) and inter subband Zener tunneling (dashed line).

Fig. 2: Current versus applied voltage for three different lengths
across which the applied voltage drops, (a) 60 $\AA$, (b) 30 $\AA$,
and (b) 10 $\AA$. The results for tubes with five different
diameters are shown. Zener tunneling is negligible in (a). For a given
nanotube, the importance of Zener tunneling increases with increase
in the electric field strength, as in (b) and (c). For a
given applied voltage, the importance of Zener tunneling increases
with increase in nanotube diameter.

Fig. 3: The left and right columns are the nanotube band structure
close to the left and right contacts respectively. The central column
is the total transmission versus energy for three different
lengths (60$\AA$, 30$\AA$ and 10$\AA$) over which the voltage drops.
The current (Fig. 2) is the integral of the total transmission from
$\mu_R$ to $\mu_L$.
The Zener tunneling probability is negligible for the non crossing
subbands in the case of L=60$\AA$. For L=30$\AA$ and 10$\AA$, the
opening and closing of a transmission window due to the first non
crossing subband (solid line marked NC1) is seen. For L=10$\AA$, the
opening and closing of a transmission window due to the second non
crossing subband (dotted line marked NC2) is also seen. $V_a=$2.5V.

Fig. 4: The total transmission versus energy for $V_a=$3.5V.
The total transmission is equal to one in energy ranges near
the band centers in the left and right contacts, where crossing
subbands are absent at either the left or right ends. The total
transmission in between is approximately two, corresponding to 
transmission of both crossing subbands.

Fig. 5: Current versus applied voltage for a (10,10) nanotube in
the presence of defects. For $V_a < 3.1$V, the differential conductance
decreases with increase in the defect scattering strength. For $V_a >
3.1$V, the differential conductance with defects is larger than the
defects free case. This is because inter subband scattering opens
channels for transport involving the non crossing subbands. The strength
of defect scattering increases with increase in $\epsilon_{random}$.
Inset: Total transmission versus energy with and without defect
scattering, when $V_a=$4V, Note that in comparison to the defects free
case, the total transmission is larger than one in energy
windows near 0 and $-V_a$.

\begin{figure}[h]
\centerline{\psfig{file=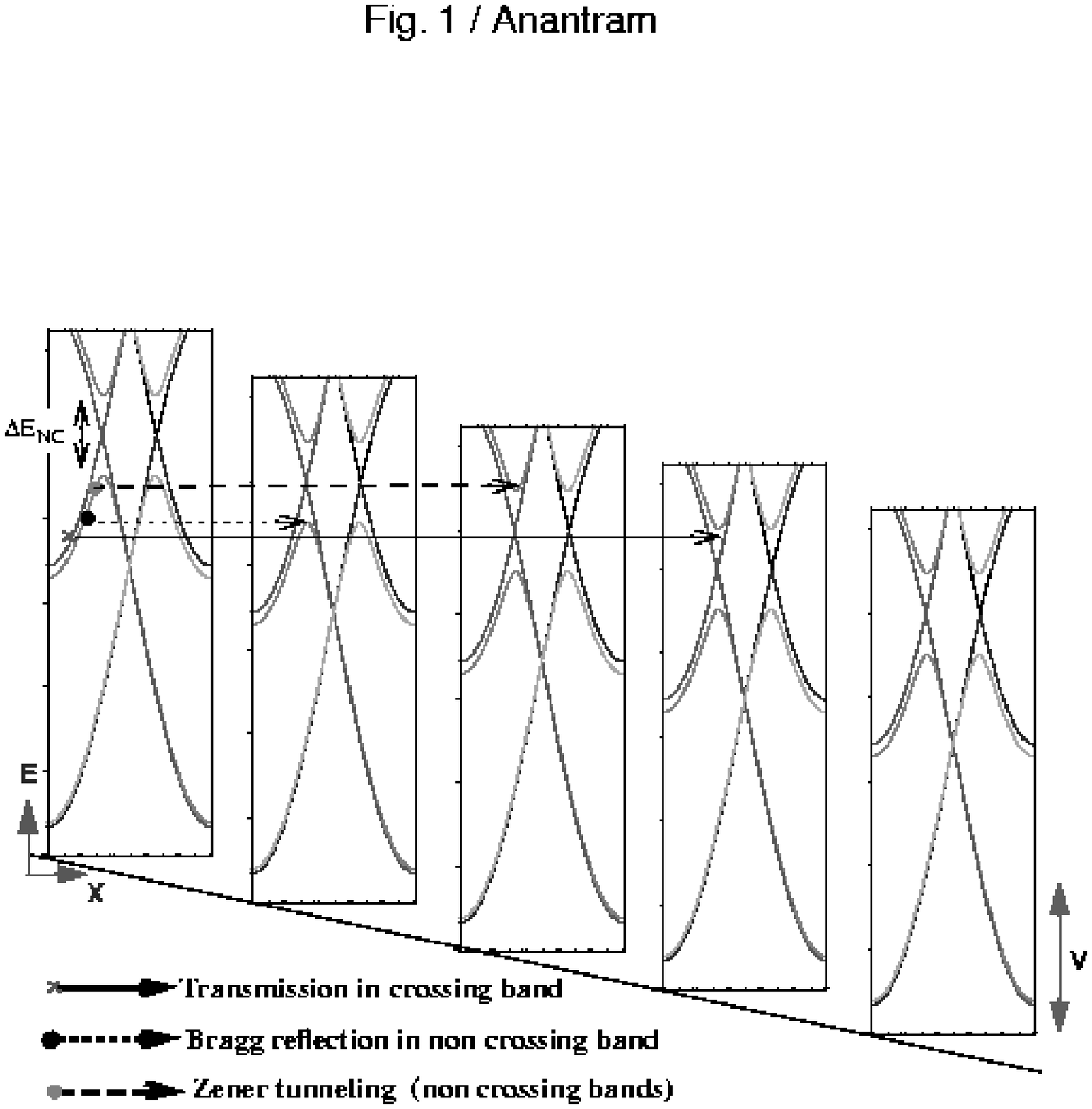,width=5.0in}}
\small
\end{figure}

\pagebreak

\begin{figure}[h]
\centerline{\psfig{file=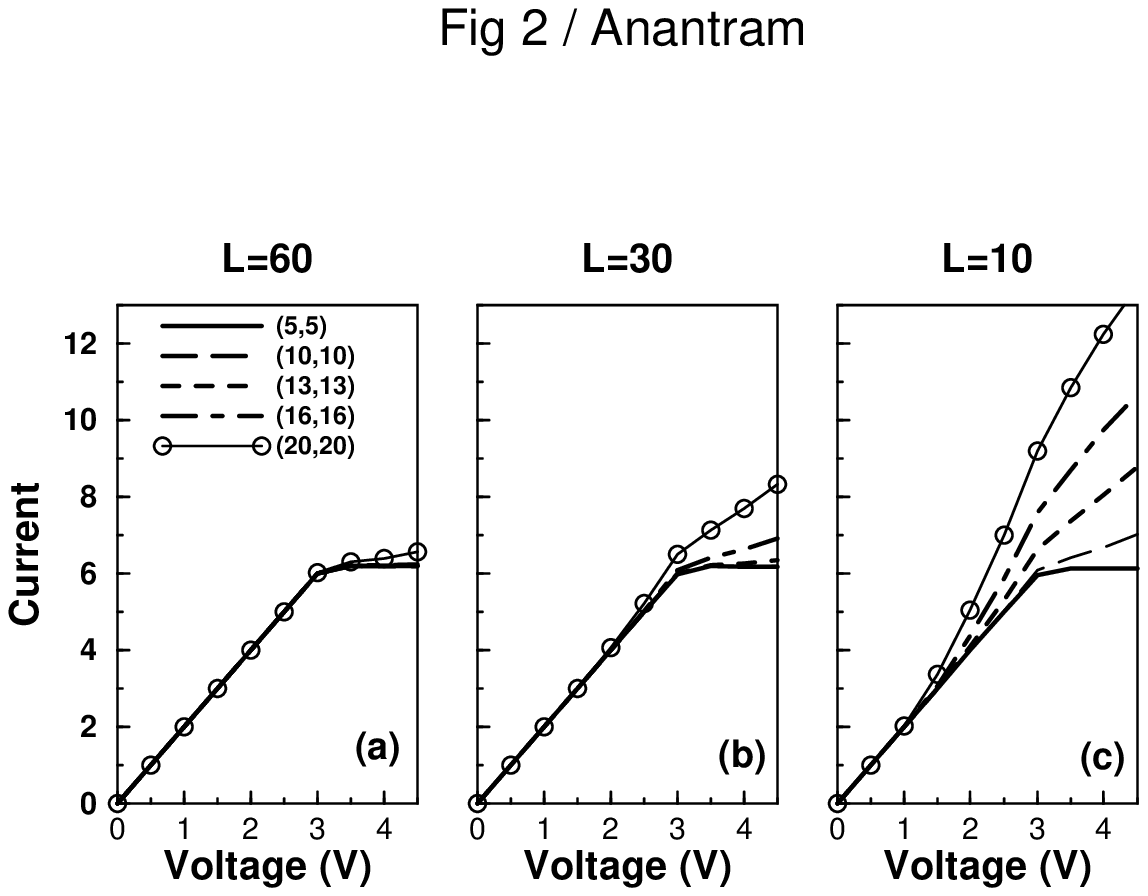,width=5.0in}}
\small
\end{figure}

\pagebreak

\begin{figure}[h]
\centerline{\psfig{file=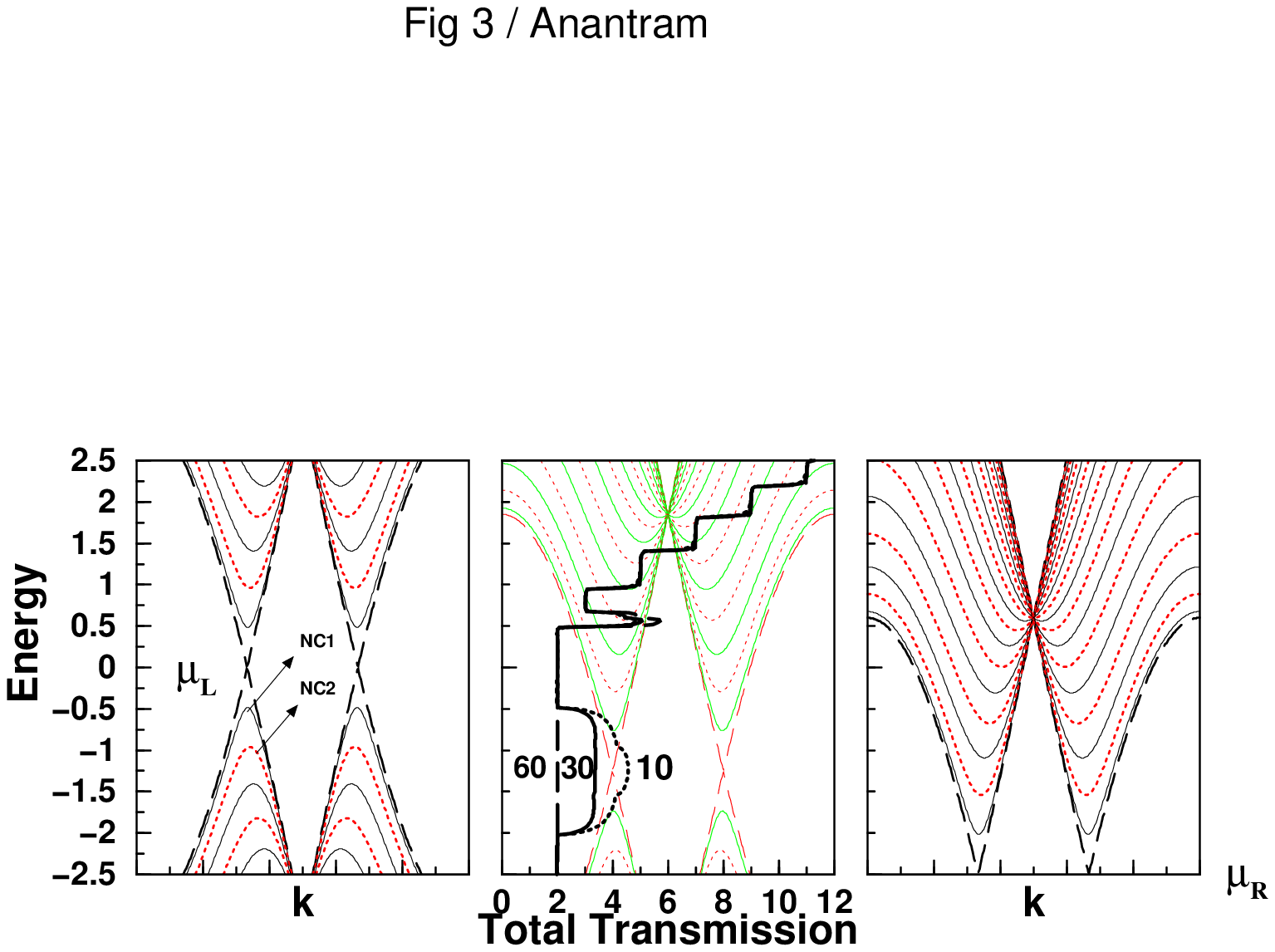,width=5.5in}}
\small
\end{figure}

\pagebreak

\begin{figure}[h]
\centerline{\psfig{file=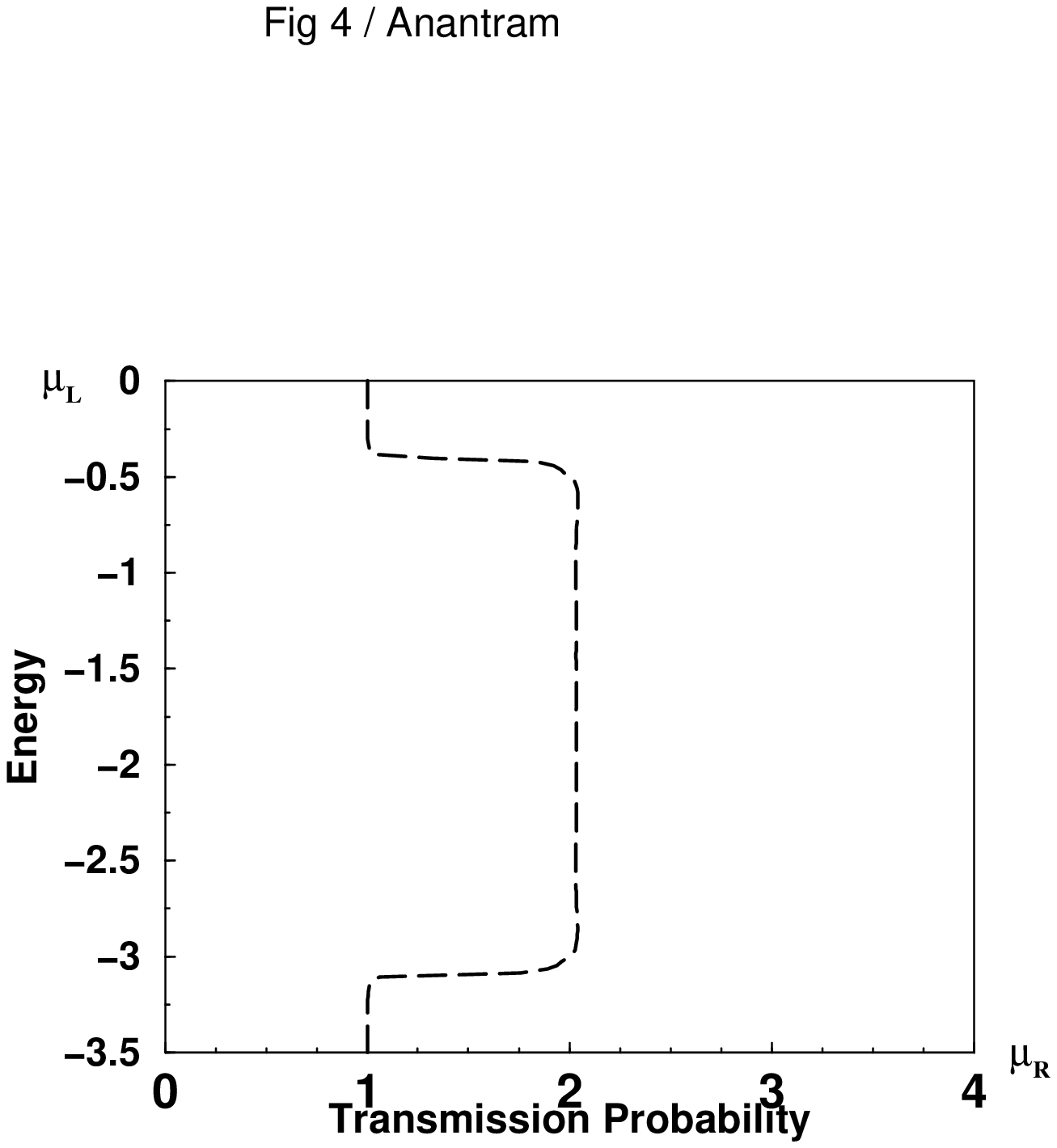,width=4.0in}}
\small
\end{figure}

\pagebreak

\begin{figure}[h]
\centerline{\psfig{file=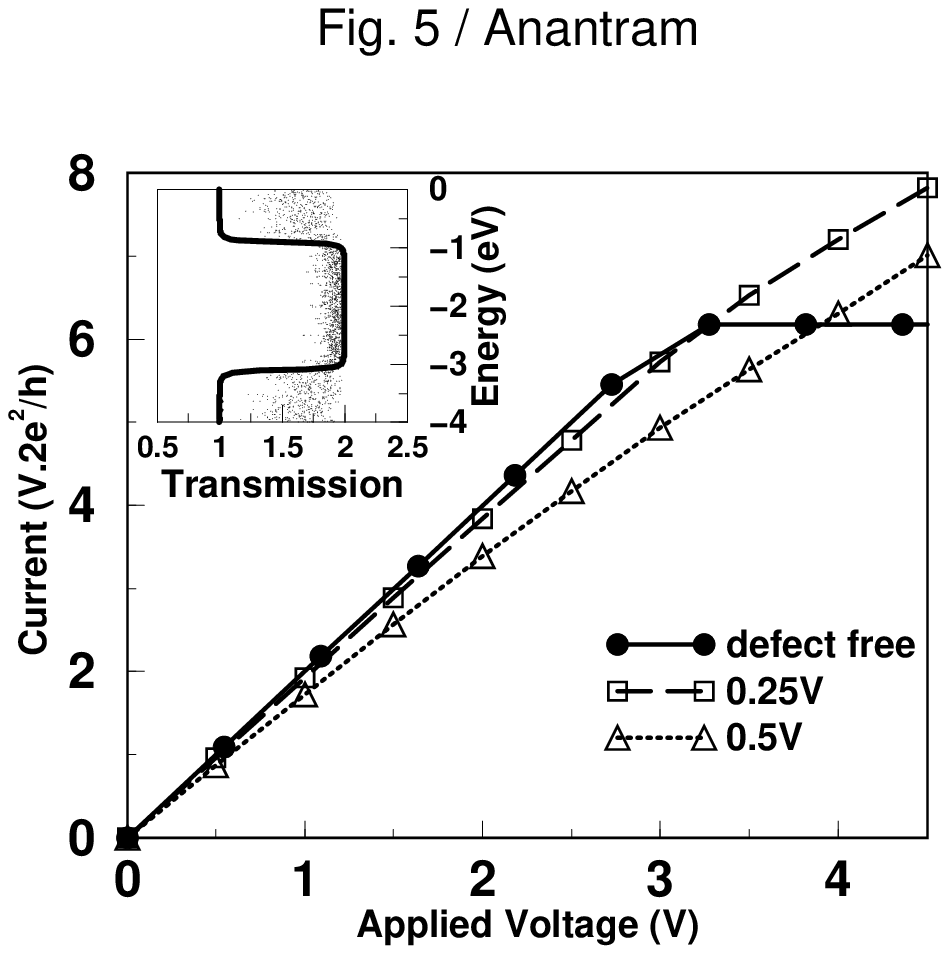,width=4.0in}}
\small
\end{figure}

\end{document}